# Temperature effects on MIPsin theBGO calorimetersof DAMPE[①]

Yuan-Peng Wang (王远鹏)[1,2], Si-Cheng Wen (文思成)[1,2], Wei Jiang (蒋维)[1,2], Chuan Yue (岳川)[1,2], Zhi-Yong Zhang (张志永)[3], Yi-Feng Wei (魏逸丰)[3], Yun-LongZhang (张云龙)[3], Jing-Jing Zang (藏京京)[1②], JianWu (伍健)[1]

[1] Key Laboratory of Dark Matter and Space Astronomy, Purple Mountain Observatory, Nanjing 210008, China

[2] University of Chinese Academy of Sciences, Yuquan Road 19, Beijing 100049, China

[3] State Key Laboratory of Particle Detection and Electronics, University of Science and Technology of China, Hefei 230026, China

**Abstract** In this paper, we present a study of temperature effects on BGO calorimeters using proton MIPs collected in the first year of operation of DAMPE. By directly comparing MIP calibration constants used by the DAMPE data production pipe line, we findan experimental relation between the temperature and signal amplitudes of each BGO bar: a general deviation of $-1.162\%/\mathrm{°C}$, and $-0.47\%/\mathrm{°C}$ to $-1.60\%/\mathrm{°C}$ statistically for each detector element. During 2016, DAMPE's temperature changed by ~8°C due to solar elevation angle, and the corresponding energy scale bias is about 9%. By frequentMIP calibration operation, this kind of bias is eliminated to an acceptable value.

## 1 Introduction

The DArk Matter Particle Explorer (DAMPE) is an orbital telescope with highresolution and wide energy band aimingat detecting cosmic rays and gamma-rays of 0.5GeV – 100TeV[1][2][3]. Apartfrom the supporting structures of the satellite, the telescope itself consists of four sub-detectors.From top to bottom (along the direction towards the center of the Earth) are plastic scintillator detectors (PSD, 82 units), a silicon tungsten tracker (STK, 768 units), bismuth germinate oxide detectors (BGO, 308 units), and neutron detectors (NUD, 4 units),where the number of units means the number of corresponding crystals, for example, there are 308 BGO crystal bars on DAMPE. Each of the four sub-detectors has its own targets: PSD for charge, BGO for energy/PID, STKfor direction/charge, and NUD for particle identification. Together, they make DAMPE an orbital telescope with the ability to measure high-energy particles up to 100TeV with good angularresolution [1].

To measure the above parameters, DAMPE has to deal with thousands of electronic channels that connect the PMTs of physical detectors with the data processor to store the informationwanted.Calibration is required to convert the charge excited by interaction between particles and DAMPE to the energy of the particles.A detailed description of the procedure can be found elsewhere[4], and only part of it is described here: the effects of temperature on the response toMIPs of the BGO calorimeters of DAMPE.

MIPs, short for minimum-ionization particles, whose energy loss in material can be quantified by the Bethe-Block formula, are the kind of particles that can penetrate materials while depositing energy only by ionizationrather than suffering nuclear reactions. This means thatthe energy they losemakes up nearly a fixed percentage of their total energy.Assisted by simulation with the same structure and (nearly)the same energy spectrum (Section 2 below introduces it briefly), the exact energy of a MIP is acquired, andis then utilized to calibrate the BGO and PSD by matching that energy with the magnitude of the ADC (AnalogtoDigital Conversion[③])signal collected bythe PMTs when a MIP hitsthem[5].A thorough understanding of the particles

---

[①] The DAMPE mission was founded by National Key Program for Research and Development (No. 2016YFA0400200). This work is also supported in part by the strategic priority science and technology projects in space science of the Chinese Academy of Sciences (No. XDA04040000) and by NSFC under grants No. 11303105 and No.11673021.

2009 Chinese Physical Society and the Institute of High Energy Physics of the Chinese Academy of Sciences and the Institute of Modern Physics of the Chinese Academy of Sciences and IOP Publishing Ltd

[②] E-mail: zangjj@pmo.ac.cn

[③] In many other cases one may expect "Convertor" instead of "Conversion". Here we use the latter because "ADC" hereafter is used as a unit measuring the charge rather than the convertor itself..

hitting DAMPErequires much more effort than MIP calibration("MIP calibration" is used hereafter for "the procedure of calibrating the ADC values of MIPs to acquire their ADCs for future use") alone. For example, the relation between different dynodes of a PMTshould also be calibrated, which can also be found in Ref.[5].

However, the magnitude of charge read from a PMT when a particle hits BGO is actually determined not only by the nature of that particle but also by the PMT itself[6]. So the environment of the PMTs will affect their behavior.Temperature is one of the most important factors which is affected by the angle of the satellite, as it gets heated by sunshine. Aprevious experimental estimation [7]tells us that its influence on the behavior of PMTs is approximately $-1\%/°C$,but the precision of that estimation wasfar from acceptable. So this paper aims at a more accurate evaluation of this effect by comparing the data from DAMPE fromabout one year, covering a range of temperature of $2 - 10\ °C$. After the comparison, a linear fit is applied, allowing more precise evaluation ofthis effect.

Simulation is usedin many partsof our analysis, and this paper introduces it briefly in the next section, followed byour procedurefor selecting MIP samplesanda brief fitting result. Afterthat is the major topic, the temperature effects of MIPs, which is arrangedin three parts: the variation of temperature and MIP results on different dates, the global relation between temperature and MIP calibration constants, andfinally the results of each BGO calorimeter. Finally we come to the conclusion.

## 2    The DAMPE simulation

A detailed descriptionof simulation in DAMPE is in preparation, and here only a brief introduction is given,withspecial attention paid to MIPs.

DAMPE uses GEANT4 as thebasis for simulation[8][9]. First we take the spectrum from AMS02 as the input of GEANT4. Now the energy of MIPs is affected by the angular distribution, so DAMPE's orbit must be taken. A techniquecalled "back-tracing" is therefore applied to target the particles according to the orbital parameters of DAMPE (time and position). This technique enables a particle in GEANT4 to swirl in a magnetic field according to the geomagnetic field at the same coordinates as DAMPE when it is being considered, so we can then re-model the spectrum for DAMPE.Because the tracks of particles coming from earth will eventually collide with the earth if we reverse-extend them, abandoning these particles offers us a fine spectrum with orbital information of DAMPE. The difference between the real spectrum and that from our simulation is quite acceptable, as can be seen from Ref. [10].

With simulation data, we can find perfect MIP samples whose exact energy is known from input. Using a digitization technique, the ADCs of the MIPs are also provided. These values are used as defaults, for example in the case of a new environment where no calibration for MIPsis available, because they are independent of real data. Later in the following analysis, they are also used as a reference to give a normalized relation where the real values are not concerned.

## 3    Procedureforselecting MIPs

MIPsare defined, in this work, as particles which loseenergy only by electromagnetic reactions with materials.Consequently, the ratio of MIPs when penetrating materials is determined by thegeometric structure and thematerials themselves.Apart from ionization energy loss, MIPs hardly interact with materials, making their track quite straight and clear when penetratingDAMPE. These factspave the way for determining whether a particle can contribute to MIP calibration.Five different filters for MIPshave beendeveloped accordingly and are listed below.Of these, onespecific filter is designed only for DAMPE. All filters below are listed in the order they are arranged in the code. If one filter rejects a particle when it is being checked, it will be thrown away immediately without checkingthe rest of the filters.Reference[11] details the software framework we use.

- Trigger mode

Exposed directly to cosmic rays, DAMPE gets hit by millions of particles every minute, the majority of which are low-energy and thus low priority. This makes the data selectionprocedure extremely important.The trigger mode of DAMPE decides what to record. Severaltrigger modes are designed for DAMPE because it has many scientific tasks.Some modesare designed to increase the number of MIPs due to their significance in calibration. In this way, selecting certain trigger modes can help preclude a large number of particles when only MIPsareneeded.MIPmodes are enabled only at low latitudes (20°S − 20°N). Details ofthe DAMPE trigger systemare available elsewhere.This is a DAMPE-specific filter.

- Penetration

This filter, making sure thata particle hasdeposited energy inboth the top and bottom ofthe BGO calorimeter, aims at selecting MIPsthat penetrate DAMPE thoroughly by throwing away those entering or leavingDAMPE through one side ofthis satellite.It is currently a compromise to sacrifice the efficiency of MIP selection for the sake of energy estimation, becauseotherwisethe ratio of the energy from the BGO calorimeter to that of the particle cannot be easily decided, making it impossible to acquire a good evaluation of the energy of the particle. Now that the energy spectrum of MIP is necessary when calibrating, these "crippled" MIPs are abandoned[①].

- Total energy

The proportion of its total energy that a MIP deposits in material during interaction is approximately fixed, although that ratio varies with the energy of the particle. This gives us a third filter to abandon particles that have deposited more than 40 times the energy of a typical MIP (or 21.8MeV).

- Number of hits per layer

MIPs interact with materials in a simple way without breaking apart, so they hardly interact with BGO units they don't penetrate. This is the basis for the fourth filter: if too many detectors on a plane respond to a particle, it is probably not a MIP. To avoid noise, only the units the particle penetrates and those adjacent are considered, and a unit is considered to have responded only when more than 0.2 times the typical MIP energy (4.36MeV) has been deposited.

- $\chi^2$ of the track

A MIP penetrates materials in a simple way, so it must follow a clean track, which means that the $\chi^2$ of the track is expected not to be very large. This gives us our fifth filter. In this filter the track is calculated in a rough way by linearly fitting the energy-weighted coordinates of the responding BGO unit and the two next to it on each layer, and then omitting particles with too high a $\chi^2$ ($> 2$ for now, $\sim 1$ typical). Here we use energy for weighting because it helps improve the resolution. Given its total energy, the energy a MIP deposits in one unit may not exceed the noise so much that considering all units unbiasedly would bend our track far off the real one, while we can expect a higher precision if energy-weighted coordinates are used instead because they can increase the significance of units with high energy and decrease that of some noise units. Details of track reconstruction go beyond the scope of this paper.

These are the five filters developed to pick up MIP samples from DAMPE data. Approximately 30000 MIPs are selected each day from the 5 million particles recorded by DAMPE. This, however, doesn't mean the efficiency of our MIP selection is only ~0.6% (30k / 5M), because the satellite does not only measure MIPs. The denominator should be the real number of MIPs each day, which seems completely inaccessible at present. We therefore estimate the efficiency based on simulation and conclude an efficiency of 88%, which is the only result we can trust. In order to get adequate samples to fit, data from 5 continuous days is accumulated for calibration each time. This, however, assumes that a span of time does no harm to our resolution. Fortunately for us, it is observed that the temperature changes by no more than $0.25℃/day$, so this stability serves as the foundation for our method considering the temperature effect of $\sim -1\%/℃$.

Selecting MIPs is only the first step. The second step is to fit them to get an estimation of MIP energy deposit. First we use the track of a MIP to correct path length by a factor of $\cos\theta$, then a convolution of Landau and a Gaussian function is used to fit their spectrum. After the fitting, the peak of the fitting convolution is used when reconstructing the energy of particles instead of the Most Probable Value (MPV) of the Landau distribution[②], but hereafter we use "MPV" in our discussion. Whether peaks or MPVs should be used is still in debate, and currently peaks are used (MPV was chosen before in early versions). Plotted in Fig.1 is a sample fitting a MIP histogram during an instance of calibration. This shows that our procedure works well. After all, rather than individual MIPs, we care more about their spectrum, for the value we require, peak or MPV of the function, comes from statistical analysis. However good the sample looks, it is rather complicated to fit with such a convolution. Details of our fitting are provided later in Section 4, where more results are shown.

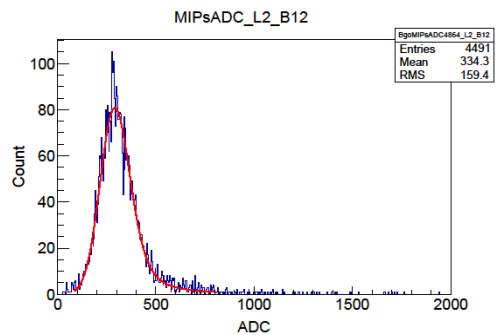

Fig.1 A sample MIP fitting histogram and its fitting function. The horizontal coordinates, marked ADC, show the charge read from the PMT, and the vertical the counts. We use a convolution of a Landau function and Gaussian function to fit the data with an empirical range, and then draw the result onto the data as the red line. The total number of counts in each histogram is ~ 4000, adequate for fitting,

---

[①] A deeper understanding of this procedure requires that a MIP penetrate the top and bottom surface of the BGO unit that is being calibrated, because we are calibrating the BGO unit instead of the whole DAMPE.
[②] Each bar has its own peaks, so we need to select the peak of the bar a particle hits.

as can be seen from this histogram, where the red function looks fine.

## 4 Results

MIPsarecalibrated every day to make up for the possible daily change of temperature, however slight, and this offers us about four hundredresultsup tonow. With thermistors and other similar electronic devicesuniformly distributed inside DAMPE, the temperature field of DAMPE can be calculated. The relation between the MIP parameters and temperature can then be studied.

The variation of averageglobal temperature of DAMPEis shown in Fig.2. This is the average of all BGO units,giving an average temperature of the field of DAMPE, averaged over of 5 days,because each calibration is done using 5 days' worth of MIPs. In Fig.2, parametersare marked as the first day of the 5 days.

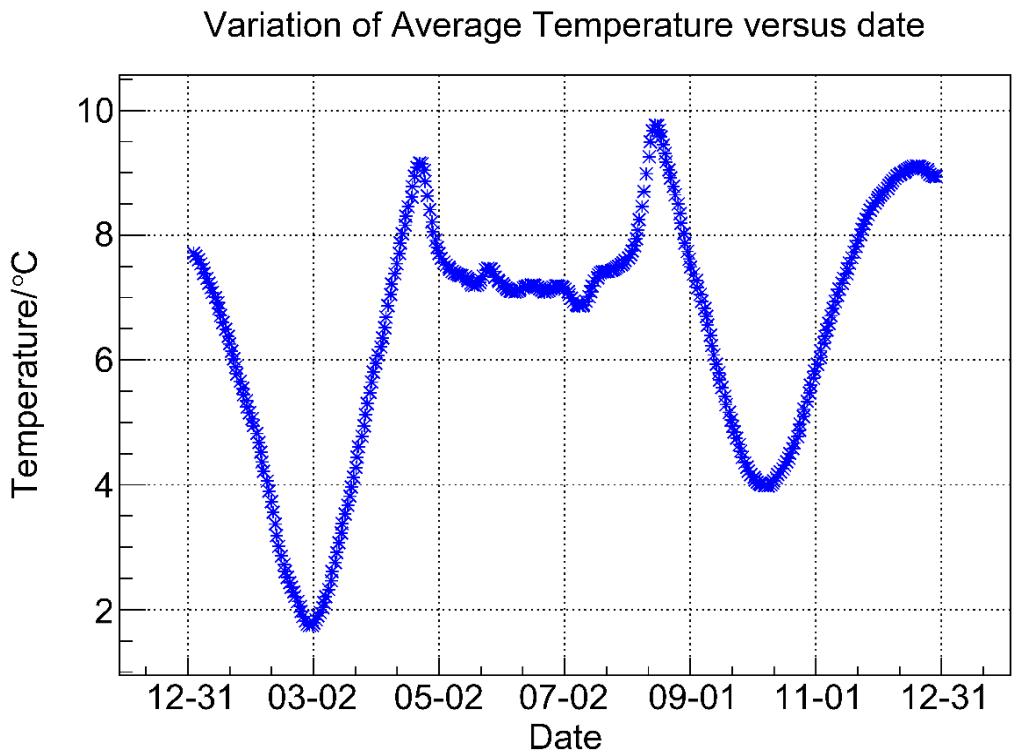

Fig.2Variation of averaged global temperature of DAMPE.It covers a range of roughly 2℃ − 10℃since launch. This tendency is largely decided by the orbit, which affects the exposure of the satellite to the sun.

Fig.2shows that the temperature changed by at most~8℃overthis first year. The temperatureis always changing but with varying trends.It kept decreasing until around the end of February, then increasedcontinually until about Apr. 25, before decreasing again until about May 2.After this it started to wobblearound 7.2℃, until the middle of August when it climbed to its peak, after which the temperature decreased until October, before finally climbing again until the end of the year. This can be explained by the direction of the satellite(as well as solar elevation angle),as this affects the efficiency with which it absorbs sunshineand heats up.The direction is decided by its orbit:DAMPE follows a solar-synchronized orbit whose inclination is about 97°, which gives DAMPE a periodic variation of the time it isexposedto the sun, giving rise to this pattern of temperature variation.

Plotted in Fig.3 is the variation of MPV of MIPs as a whole. When calibration is done, 308 MPVs are calculated for each ofthe 308 BGO units. To acquire a point on the diagram, each of those 308 MPVs are divided separately by thoseof the same unit but calculated instead using simulation samples, and 308 ratios result. Then a Gaussian function is used to fit these 308 ratios to get itsmean and sigma, which are finally plotted onto the diagram as a point and its error bar. In this way, the 308 MPVsacquired eachtime are condensed into one point onto the diagramrepresenting the global effect of the result of calibration at one time as a whole instead of one particular BGO unit. This also explains whythe error bar of any pointin the diagram is quite small: it can be taken as the sigma of 308 "repeated" measurements.The MIPsare calibrated with 5 days' worth of data each time, and to plot them, the first date was chosen, as is the case for the temperature above.

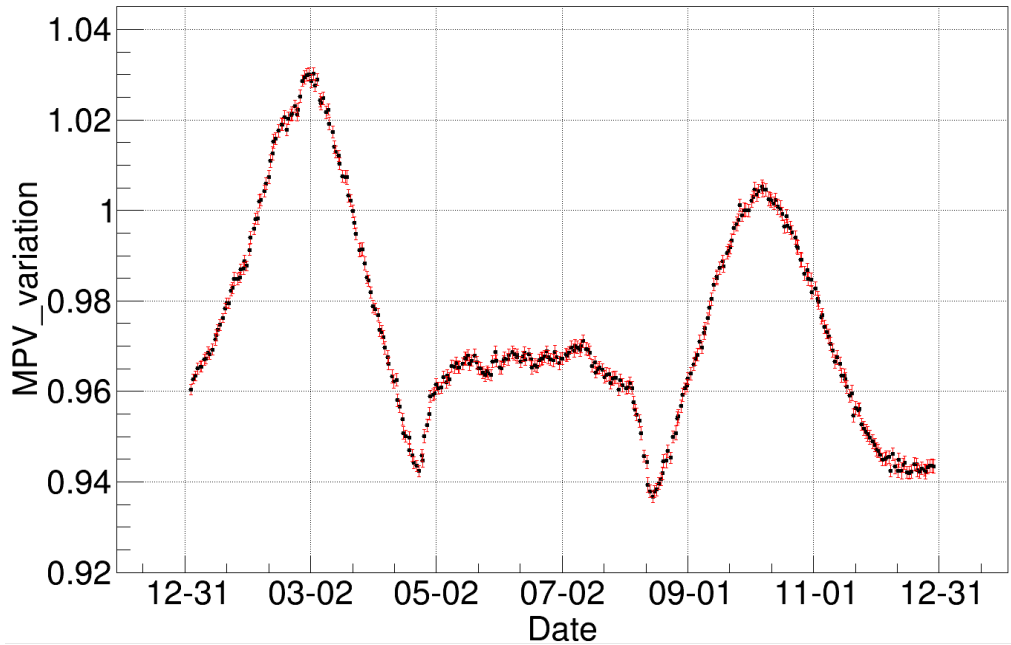

Fig.3 Variation of MIPs as a whole. The vertical coordinates represent a statistical evaluation of MIP calibration each time, and the horizontal ones are dates of calibration where a timestamp from the first day is plotted. There are some gaps in the diagram, for example the two isolated dots to the left of 09-01, but it is quite safe to ignore them (see the text).

By roughly comparing the two diagrams one can see that they match well, with minor deviations. Some gaps on Fig. 3, such as the two located to the left of 09-01, however wide they look, do not come from the unavailability of data on that date but from the way of selecting their horizontal coordinates: a timestamp from the first day is selected but in an arbitrary way. A better algorithm would help improve the performance by, for example, smoothing some unnecessary sub-structures in the diagram, and it is currently being developed.

The effect of global temperature MIPs is shown in Fig. 4. It is plotted using the values for each day from the two diagrams. Temperature is shown on the horizontal axis, and the MPV of the MIPs on the vertical. The standard deviation of each MPV is also plotted but is very small. The red line shows the linear fitting function whose parameters are given in the top right corner.

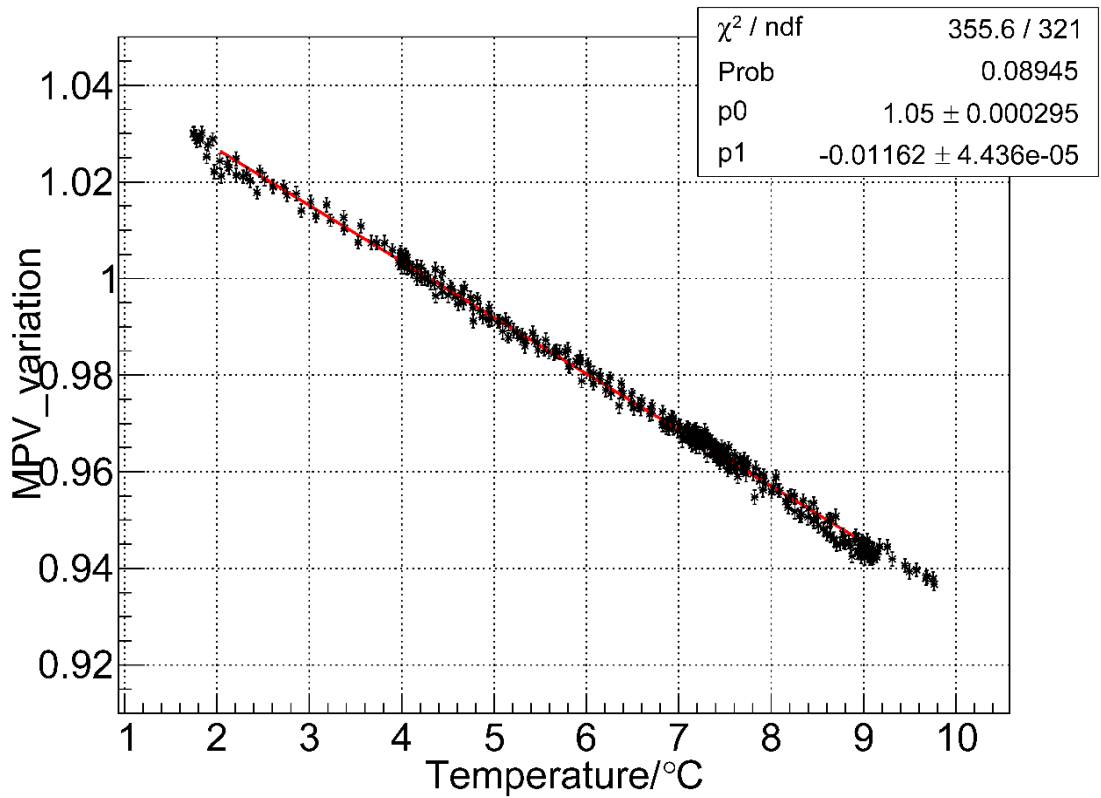

Fig.4 Variation of MPVs versus temperature, showing perfect linearity. Due to thermal imbalance on either peak, the fitting range is not as wide as the data. The cluster of points near 7℃ is due to the change of temperature of the satellite, as can be seen above in temperatureplot in Fig.2.

From Fig.4 we can see a perfect linearity between the temperature of DAMPE and the MPV of the MIPs. It gives a deviation of $-1.162\%$/℃ globally, that is to say, a temperature change of 1℃ brings about a global deviation of MPV of $-1.162\%$. The fitting range of this diagram is only $2-9$℃ instead of the whole data range, because when it is too hot or too cold, the thermal equilibrium hasn't been reached, making it difficult or even impossible to estimate the temperature field.

The final result is the temperature effect on each BGO detector. For the sake of convenience, each of the 308 BGO units is afterwards referred to as one "BGO bar" or only "one bar"("bar" is used due to its dimensions of 600mm×25mm×25mm). Following the global case, temperature should come first. It is however omitted due to the similarity between different bars: different bars show only minor changes in the shape of the temperature variation, while the overall trend is the same as the global one concluded from Fig.2. This is easily understood in terms of thermal equilibrium. The temperature diagrams are therefore omitted here and here we start with the MIP ADCs. Results for 4 bars are shown in Fig. 5. The different bars have been normalized (so the peak of each diagram is 1, and errors are scaled accordingly) so that one can focus on the trends rather than the absolute values. This is because the absolute values of each diagram vary from 200 to 700, which makes it inconvenient to evaluate the trends. The density of points makes it difficult to plot them all on the same figure, so there are four individual diagrams. We will come back to real ADCs when linearity is concerned. Also, the MPV from the fitting function of the ADCs is plotted each time, and vertical axes are thus labeled accordingly.

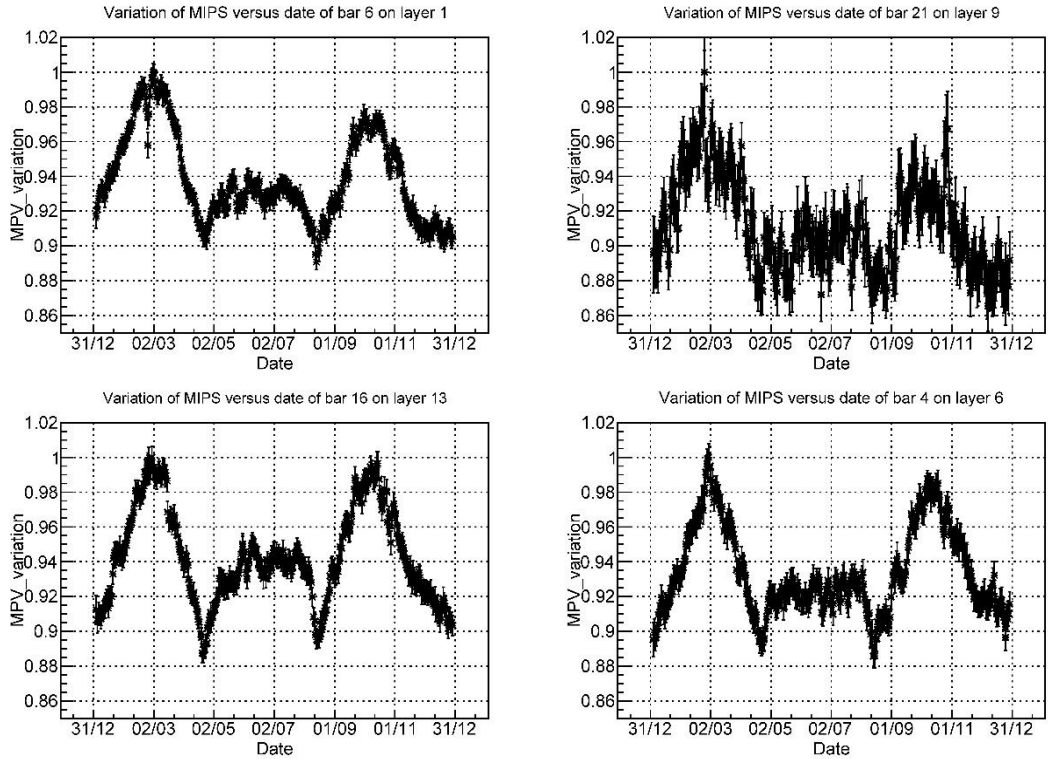

Fig.5 Variation of 4 selected bars. Vertical axes are all normalized to the same scale with 1 the largest, to allow clearer comparison of the shapes. Deviation here is quite large compared to temperature because fitting errors, statistics, and initial values of fitting data of one bar affect the result more than in an averaged analysis. Shapes roughly match those of temperature plots. Bar numbers are given in the title of each histogram.

Here the MPVs of each bar fluctuate more than in the global case. This can be understood by the fitting procedure. Fitting relies on the initial values of each parameter. This dependency dominates in the case of the convolution of a Landau and Gaussian function, where a minor deviation in the initial value may lead to visible mismatch, and MIP calibration depends on this convolution. Initially, recursive efforts were made to improve the initial conditions as well as the fitting range for better results, then a two-phase fitting process was developed, where the first trial aims only at providing the initial conditions for the real fitting in the second phase. In doing this, only the MPV of the first trial is used to construct the initial conditions for the second fitting. We use only MPV, which can be roughly estimated as the peak of the data, because not too many steps are needed to acquire this parameter, because of efficiency, and because it is observed that the MPV is still reliable even if the fitting fails. This two-phase fitting method recovers many failures from the previous fitting method because the new initial conditions it uses are more reasonable. If the reduced $\chi^2 (= \chi^2/NDF)$ is larger than 3, we consider this two-phase fitting to have failed as well, and then a third fit is performed. For this third trial, the MPV of the second fit is still consulted to give the initial conditions but in a slightly different way than that in the second trial to avoid further failure, and these differences also come from our recursive trial. In other words, each histogram is fitted at least twice but at most three times, and these efforts give us the smoothness in fitting functions like Fig.1 with few exceptions in the end. The global relation stated above is concluded by a statistical study of 308 bars where individual deviations are largely tolerated.

The final part is linearity of temperature and MIPs, shown for four bars in Fig.6. A statistical analysis for all the bars is shown in Fig.7 and 8.

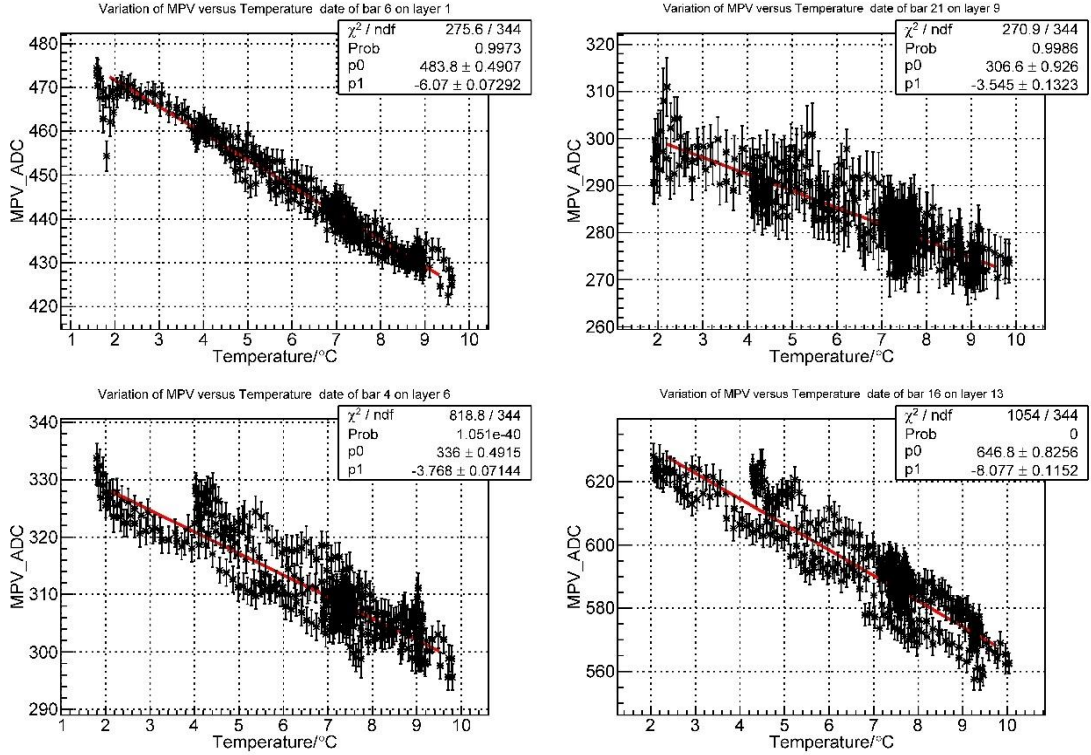

Fig.6 Relation between temperature and MIP constants of the same 4 BGO bars as chosen above. The top two diagrams show good linearity while the bottom two are less good. The relation here deviates from linearity due to the relatively large deviation of MPVs of MIPs of each bar as displayed and explained in the text.It remains true, however,that higher temperaturebrings about lower MIPs.As in the global case, not all the data points are used, and the ranges here correspond with 2 – 9℃ in the global case (see the text).

Following therestricted range, the first step is find the corresponding range for each unit. 2 - 9℃is not suitable here because thermal equilibrium doesnot mean that all bars share the same temperature. It is possible to calculate using linear interpolation, provided that the same tendency holds for all bars with minor deviation, as mentioned before. It is an easy method with good resolution, much easier than the most reliable way which requires a thorough calculation for either condition. From these figures one can also obtaina standard for selecting bars: normal bars with good linearity (such as the Fig. 6(a) and (b), ~ 83% in all bars), and abnormal bars with bad linearity (such as theFig. 6(c) and (d) ~ 17%) where good linearity means a bar with a fitting function whose$\chi^2/NDF < 1.19$ (this "1.19" is actually an empirical standard which we set by using a typical ambiguous bar from early analysis). A closer look at those abnormal bars gives another hint: they are usually located on or near the edge of the308BGO units of DAMPE[①].It may explain their behaviorto understand the lack of statistics compared with the rest.However abnormal a bar appears, their slopes are always negative.

Fig.7and 8show the distribution of fitting parameters, where a "ref" in the title (short for "reference) means that the values inserted have been divided by those from simulation.

---

[①] All 308 BGO bars are arranged in 14 layers of 22 BGO bars each.The index of bars, either of its layer or its bar, starts with 0, for example, layer 0 bar 0 is the first bar, layer 8 bar 10 gives the 11[th] bar on layer 9, and so on.

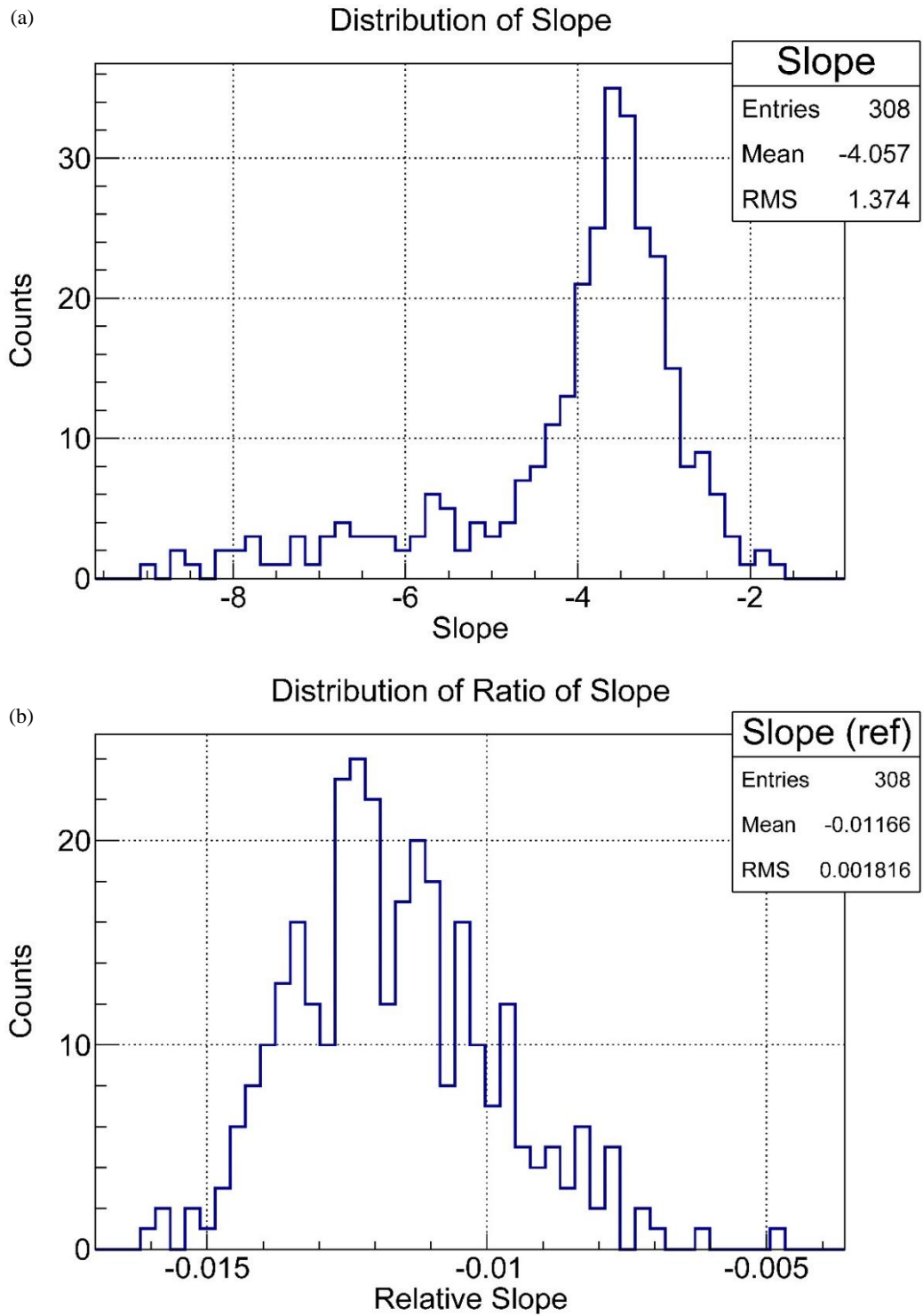

Fig.7 Distribution of slopes of linearity between temperature and MIP constants of 308 BGO bars. All slopes are negative. (a) The absolute values, calculated directly by linearly fitting the MIP constants and temperature instead of a ratio. (b) The relative values, or ratio of the value of each bar to that acquired by simulation results. The average ratio is -1.16%, which matches the global result well..

  The absolute value of the average of slopes in Fig.7(a), $-4.057$, is very large compared to the other figure where values have been divided. This direct insertion of slope looks quite loose and is less statistically important than Fig. 7(b). The values in Fig.7(b) have been divided by simulation results. It gives a mean relative slope of -0.01166, indicating an average temperature effect of about $-1.166\%/℃$, in accordance with the global result -1.162% presented above. The behavior of different BGO bars varies from $-0.47\%/℃$ to $-1.60\%/℃$. If this effect were ignored, the energy resolution would definitely suffer: considering the temperature change of about 8℃, ignoring temperature would induce a deviation in energy of about 9%.

For completeness, the distribution of intercepts is displayed in Fig.8. The mean of the relative intercept, 1.051, is in accordance with the global mean in Fig.4.

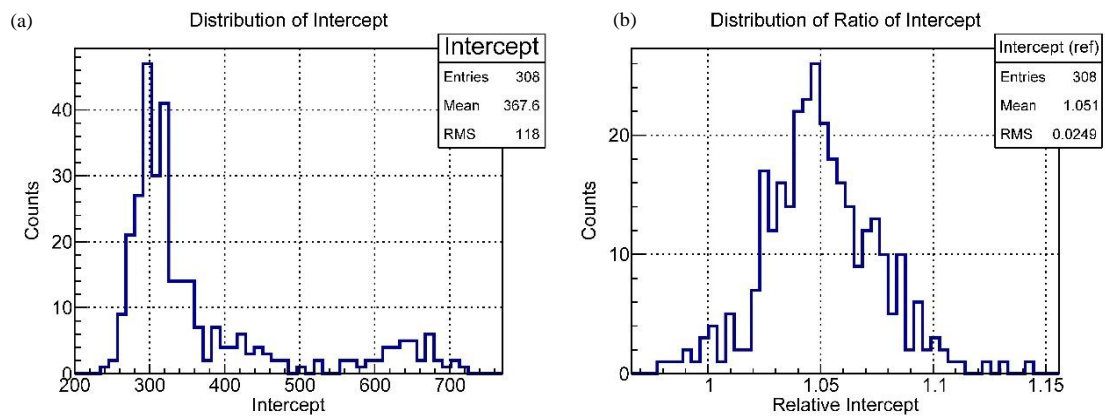

Fig.8 Distribution of intercepts.

## 5  Conclusion

Data from DAMPE has been used to calibrate MIPs for its BGO detectors. Thermistors uniformly installed in DAMPE enable us to estimate its temperature field. Combination of the calibration results with the temperature field makes it possible to analyze the temperature effects on MIPs of BGO bars (as well as PMTs).

This analysis has been done not only on DAMPE as a whole to analyze the global behavior, but also on each of its 308 BGO bars. The global analysis gives a temperature effect of $-1.162\%/℃$, which means that every degree Celsius brings about a global deviation of -1.162% to the behavior of DAMPE's BGO calorimeters. In terms of one bar alone, the average of all 308 bars gives -1.166%, and individual behavior differs from one bar to another from $-0.47\%/℃$ to $-1.60\%/℃$. From Fig.2, there have been changes in temperature of $\sim 8℃$ since the launch of DAMPE, which would have introduced a global energy bias of ~9% on DAMPE if the temperature effect was ignored, and would be even worse for individual BGO bars.

*Acknowledge:*

The authors sincerely thank the whole DAMPE collaboration, without whom it would be completely impossible to perform this analysis. We especially thank Prof Jin Chang, the chief scientist of DAMPE, who established this project and had gathered this team. We also thank Prof Ming-Sheng Cai and his DAMPE payload group for offering us an excellent detector system. We also thank the Scientific Application System for providing us with reliably reconstructed flight data. All diagrams have been plotted with ROOT (https://root.cern.ch/).